# Atomic-Scale Structure and Chemistry of YIG/GGG Interface

*Mengchao Liu[1], Lichuan Jin[2], Jingmin Zhang[1], Qinghui yang[2], Huaiwu Zhang[2], Peng Gao[1,3,4,a), Dapeng Yu[4,5,6]*

[1]Electron Microscopy Laboratory, School of Physics, Peking University, Beijing, 100871, China
[2]State Key Laboratory of Electronic Thin Films and Integrated Devices, University of Electronic Science and Technology of China, Chengdu 610054, China
[3]International Center for Quantum Materials, School of Physics, Peking University, Beijing, 100871, China
[4]Collaborative Innovation Centre of Quantum Matter, Beijing 100871, China.
[5]State Key Laboratory for Mesoscopic Physics, School of Physics, Peking University, Beijing 100871, People's Republic of China
[6]Institute for Quantum Science and Engineering and Department of Physics, South University of Science and Technology of China, Shenzhen 518055, People's Republic of China
Authors to whom correspondence should be addressed: a)p-gao@pku.edu.cn;

**Abstract:**

$Y_3Fe_5O_{12}$ (YIG) is a promising candidate for spin wave devices. In the thin film devices, the interface between YIG and substrate may play important roles in determining the device properties. Here, we use spherical aberration-corrected scanning electron microscopy and spectroscopy to study the atomic arrangement, chemistry and electronic structure of the YIG/$Gd_3Ga_5O_{12}$ (GGG) interface. We find that the chemical bonding of the interface is FeO-GdGaO and the interface remains sharp in both atomic and electronic structures. These results provide necessary information for understanding the properties of interface and also for atomistic calculation.

Spin waves (magnons) that have a large group velocity up to a few tens of μm/ns and a frequency in the gigahertz/terahertz range,[1-4] are promising for the application of information transport and processing,[5-10] as the conventional semiconductor devices are approaching their limitation.[11] One promising candidate material for the spin-wave devices[14-16] is yttrium iron garnet ($Y_3Fe_5O_{12}$, YIG),[12-20] which has the smallest relaxation parameter, high Curie temperature, excellent chemical stability[5, 21-23] and a very low damping coefficient and thus allows the magnons to propagate over several centimeters in distance.[24-28] For the large scale magnonic circuits integration, YIG is usually required to be in the form of thin film with smooth interface and thickness in nanometer scale in order to be compatible with conventional silicon technology.[16, 28-30] In fact, the energy consumption can also be effectively reduced in the thin film YIG devices.[13, 31] Particularly, the nanometer-thick YIG film is highly desirable for construction of spin wave nonreciprocity logic devices and voltage switched magnetism. However, when the thickness of YIG film decreases, the effects of interface between YIG and the substrate are expected to become pronounced or even may completely dominate the properties of the entire devices. Therefore, it's of great significance to study the atomic structure, chemistry and electronic structure of interface of YIG thin film.

In this paper, we employ aberration-corrected scanning transmission electron microscopy (AC-STEM) and spectroscopy to study the YIG film on the gadolinium gallium garnet ($Gd_3Ga_5O_{12}$, GGG) substrate.[27] The recent advancements of AC-STEM imaging enable us to directly visualize the atomic bonding at the interface. In addition, combining atomically resolved imaging and spectroscopy such as energy-dispersive X-ray spectroscopy (EDS) and electron energy loss spectroscopy (EELS) in the STEM mode allows us simultaneously to determine the elemental distribution and electronic structures of the heterostructure. By combining these state-of-the-art electron microscopy and spectroscopy techniques, we reveal the interfacial bonding of YIG/GGG is FeO-GdGaO. No significant elemental diffusion is observed at the interface. The EELS measurements show that the electronic structures of interfacial Fe remain the same with that in the interior film. Such atomically sharped interface in both chemistry and electronic structures indicates it is possible to fabricate ultrathin YIG film for future nanodevices for which no intrinsic interfacial zone exists at the YIG/GGG interface. The detailed structure information also provides necessary information for future atomistic simulation of the interface.

A cross-sectional atomically resolved high angle annular dark filed (HAADF) image of YIG is presented in **Figure 1**(a) with the atomic model being overlapped. The red arrows mark the interface of YIG and GGG. Since the HAADF image is Z-contrast (atomic number) image, in which the contrast directly reflects the atomic number of the element, the darker side of the image is YIG and the brighter side is GGG. It can be noticed that O is invisible in the HAADF image. The HAADF image shows perfectly epitaxial growth and the interface is atomically sharp. The overlapped atomic model highlights the atom positions, which will be discussed below. The crystal structure of YIG is cubic with a dimension 12.376 Å in unit cell and houses 80 atoms. In each unit cell, there are twenty $Fe^{3+}$ ions occupying two different sites. Among of them, 8 $Fe^{3+}$ ions occupy octahedral sites and 12 $Fe^{3+}$ ions with opposite magnetic moment occupy tetrahedral sites.[5] YIG and GGG have the same garnet structure. The mismatch between YIG and GGG is smaller than 0.05%.[32] This makes the high quality and defect-free unstressed film fabrication possible. In addition, for the best matching, we also dope YIG by lanthanum lightly.[7] Therefore, no dislocations are observed at the interfaces for all these YIG thin films.

The atomically resolved EDS maps of YIG are shown in Figure 1(b)-(e), which are element Fe,

Y, O and Fe along with Y respectively. The atomic model on the EDS map in Figure 1(e) further highlights the locations of Fe and Y atom columns. Figure 1(f)-(i) show the distribution of elements Fe, Y, Ga, Gd of the YIG/GGG interface, the yellow arrows mark the interface and the scale bar in these figures is 1 nm. These EDS maps are acquired at the same area as shown in Figure 1(a). The yellow arrows mark the interface position based on the Z-contrast image. There are Fe atoms diffuse across the interface into GGG, while the Y, Gd and Ga remain sharp edges at the interface from the EDS mappings.

For the YIG grown on GGG (111) substrate, there are two possible interfacial bonding between them, as shown in **Figure 2**. Along the [111] direction, there are two types of atom planes of garnet structure, which we call A and B atom plane respectively (see the details in the supporting information). B atom plane in YIG (GGG) consists of Fe, Y (Ga, Gd) and O atoms while A atom plane in YIG (GGG) consists of Fe (Ga) and O atoms only. The atom planes arrange in ABAB… order inside the crystal. Therefore, the interfacial bonding should be either FeO-GdGaO or YFeO-GaO. Based on the atomically resolved EDS maps, the bonding at the interface of YIG/GGG is identified to be A/B type, i.e., FeO-GdGaO bonding. The schematic illustration of interfacial bonding is overlaid with HAADF image in Figure 1(a). The detailed structure information of the interface viewing from another two zone axis directions is included in the supporting information.

The counts of elemental distribution from the EDS maps are averaged along the interface and depicted in Figure 2(e), which shows the width of the interfacial region is ~1.4 nm. The counts of Fe near the interface is higher than those of Y compared to that in the interior film, due to the interfacial bonding of FeO-GGG and slight Fe diffusion. We measured 18 EDS maps from different locations in different TEM specimens, and the frequency distribution histogram shown in Figure 2(f) indicates the width of the transition area is equal to the width of 1.9 unit (2.3 nm). However, it should be noted that the practical interfacial region should be even thinner due to the presence of delocalization effects from the EDS measurement. Therefore, we conclude that no significant interdiffusion takes place at the interface.

To reveal the local electronic structure of the YIG/GGG interface, core-loss electron energy loss spectroscopy (EELS) experiments are carried out on the Titan Cubed Themis G2 300 aberration-corrected transmission electron microscope with the Gatan Enfinium$^{TM}$ER (Model 977) spectrometer. **Figure 3**(a) is a STEM image of the YIG/GGG interface along $[10\bar{1}]$ direction. The big green rectangle highlights the locations where the EEL spectra were recorded with a spatial step of 4.5 Å. The O-K edge and Fe-$L_{2,3}$ edge of the spectra are shown in Figure 3(b). As marked by the dashed line, the peak of Fe-$L_{2,3}$ edge does not show any detectable shift when the probe moves across the interface. Furthermore, the intensity ratio of $L_3$ to $L_2$ is sensitive to the electronic structures of Fe, too. The ratio is calculated in Figure 3(c) marked by stars which show no distinguishable change either. Since the energy of Fe-$L_{2,3}$ edge is sensitive to the Fe valence, no peak shift or ratio change indicates the interfacial Fe remains the same nature with that in the film.[32-35] The integration of $L_3$ and $L_2$ (marked by rhombus) is shown in Figure 3(c), from which we can obtain the width of the transition area is 1.8 nm, which is consistent with the EDS measurements.

The nature of the interface usually plays important roles in the properties for thin film devices. Particularly for those devices with nanometer scale, the interface properties could be dominated. By combining atomically resolved image and EDS results, we reveal that at the interface the FeO atom plane of YIG bonds with GdGaO atom plane of GGG. Slight Fe diffusion in the GGG is also observed. The EELS measurements show that the electronic structures of Fe remain unchanged at the interface. The atomically sharped interface in structure and electronic structures may indicate there are no intrinsic interfacial effects for YIG thin film devices. The finding of atomic arrangement of interface structure provides necessary information for the future atomistic simulation such as density functional theory calculations.

**Supporting Information**

Supporting Information is available from the Wiley Online Library or from the author.

**Acknowledgements**

The authors greatly acknowledge the helpful discussion from Prof. Xiaoyan Zhong and Prof. Jing Zhu from Tsinghua University, and Prof. Jia Li from Peking University. This work was supported by the National Key R&D Program of China (2016YFA0300804), National Natural Science Foundation of China (51672007, 51502007), the National Program for Thousand Young Talents of China and "2011 Program" Peking-Tsinghua-IOP Collaborative Innovation Center of Quantum Matter. The authors also acknowledge Electron Microscopy Laboratory in Peking University for the use of Cs corrected electron microscope.

**Figures and caption**

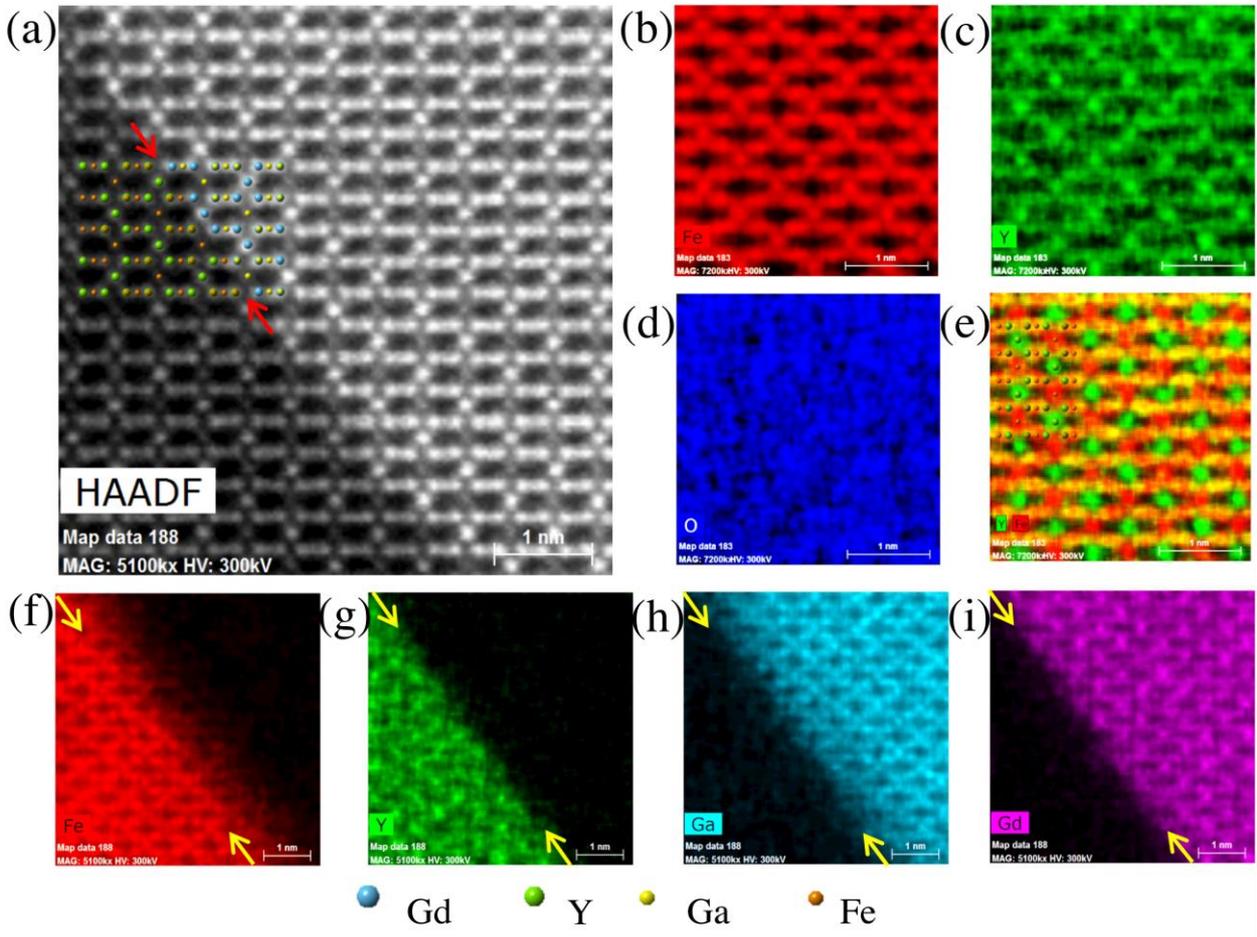

Figure 1. Atomic structure of YIG/GGG interface. (a) Atomically resolved STEM image of a YIG/GGG interface along the [10$\bar{1}$] direction. The red arrows mark the interface. The left side is YIG which appears dark contrast in the HAADF image. (b)-(e) Atomically resolved EDS maps of (b) element Fe, (c) element Y, (d) element O, and (e) overlap of element Fe and Y in YIG. The atomic arrangement model is overlapped on Figure 1(e). (f)-(i) Atomically resolved EDS maps of interface. (f) element Fe, (g) element Y, (h) element Ga and (i) element Gd. The yellow arrows mark the interface.

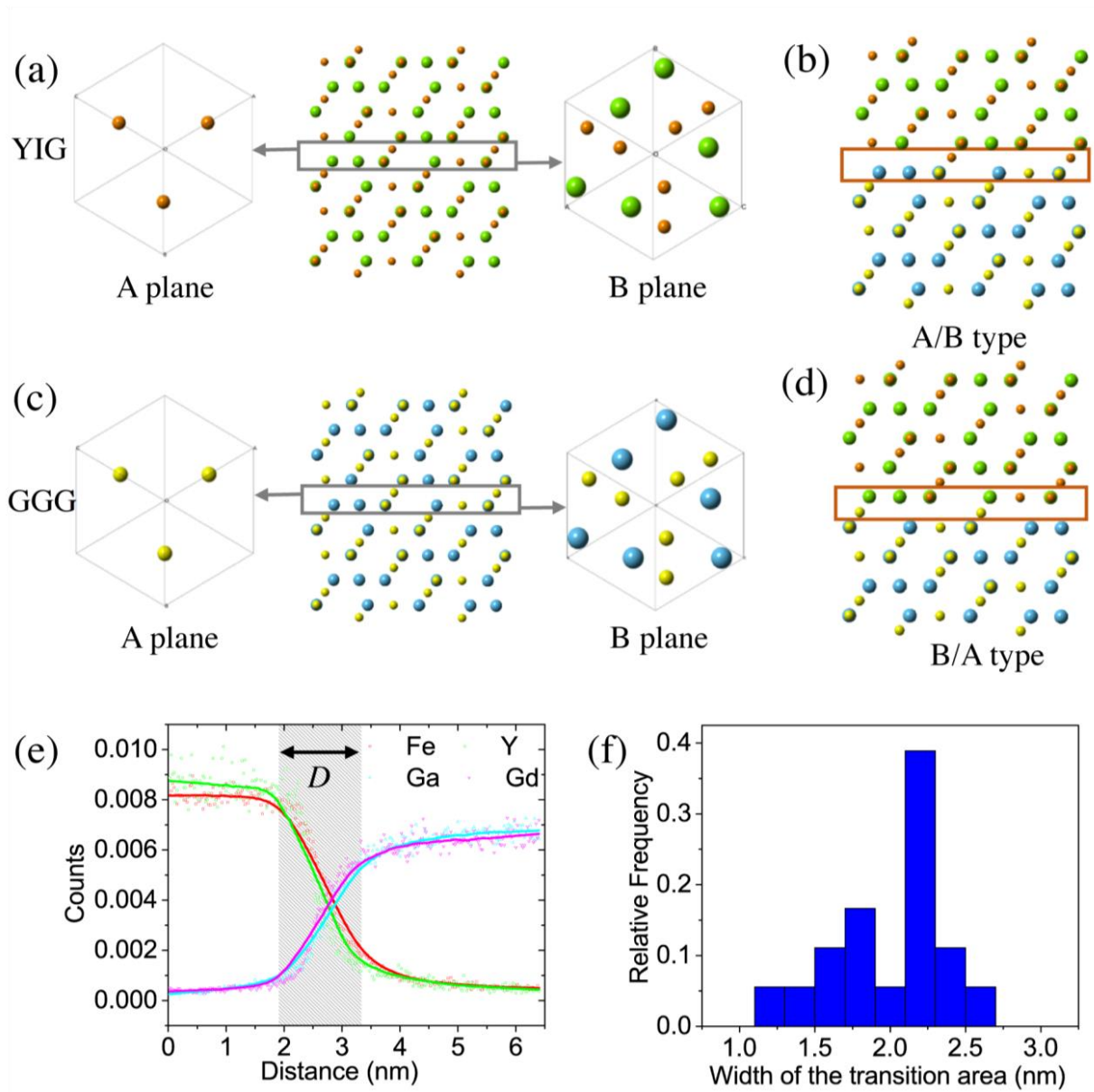

Figure 2. The atomic arrangement of the interface of YIG and GGG. (a) Two alternative atom planes of YIG along [111] direction. The oxygen is invisible for clarity. (b) A/B type bonding model at the interface between YIG and GGG. This mode is in good agreement with experimental data. (c) Two alternative atom planes of GGG along [111] direction. The oxygen is invisible for clarity. (d) B/A type bonding model at the interface between YIG and GGG. (e) The EDS results of the interface. $D$ marks the width of the transition area of the YIG/GGG interface. (f) The frequency distribution of $D$ of altogether 18 EDS results.

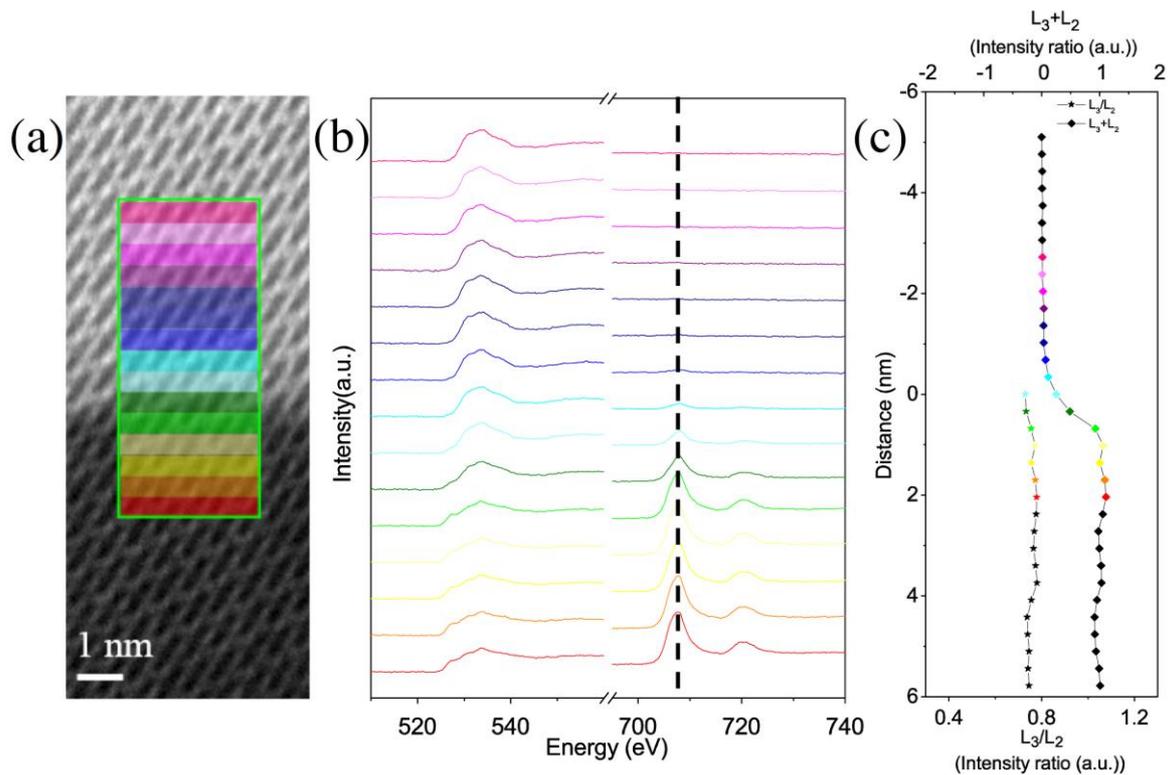

Figure 3. EELS measurements of the YIG/GGG interface. (a) A STEM image of the YIG/GGG interface along $[10\bar{1}]$ direction. The area selected to get EELS spectra is marked by the big green rectangle (consists of many small rectangles with different colors). (b) The averaged elemental line profile across the YIG/GGG interface. The spectra presented by colored lines correspond to those rectangles with the same color in (a). (c) The Fe $L_{2,3}$ white line ratio and sum across the YIG/GGG interface. The sum indicates the location of interface; while the ratio remain unchanged at the interface indicates no distinguished chemical shift in Fe at the interface.

**Supporting information**

The red arrows mark the same atom plane of two different zone axis directions which are [$\bar{3}$21] in **Figure S1**(a) and [3$\bar{1}\bar{2}$] in Figure S1(b), though the atoms surrounding them seems to be quite different from each other. It illustrates that the atom planes which consists Fe atoms only are all equivalent. Also, the atom planes which consist of Fe and Y atom are of the same kind.

Two cross sectional atomically resolved STEM images of YIG/GGG interface with different viewing directions are shown in **Figure S2**(a) and **Figure S3**(a), and the corresponding FFT patterns are presented in Figure S2(b) and Figure S3(b) respectively. The simulations of the electron diffraction in Figure S2(c) and Figure S3(c) confirm the directions of the zone axis are [$\bar{3}$21] for Figure S2(a) and [11$\bar{2}$] for Figure S3(a), respectively. The atomistic models of these two zone axes are overlapped on the STEM images. As marked by A and B plane in Figure S2(a) and Figure S3(a), it is clear that there are two kinds of atom planes which are parallel to the (111) interface. B atom plane of YIG (GGG) consists of Fe, Y (Gd, Ga) and O atoms, while A atom plane of YIG (GGG) consists of Fe (Ga) and O atoms only. The atom planes arrange in ABAB… order inside the crystal. Figure S2(d)-(g) and Figure S3(d)-(g) show the EDS results and the yellow arrows in these figures mark the interface. It can be noted that only Fe atoms diffuse across the interface for both [$\bar{3}$21] and [11$\bar{2}$] directions. This also confirms the point of view in the main text that the YIG/GGG interface belongs to A/B type as shown in Figure 2(c). The atomic structure of interface data from different zone axis directions means observing the sample from different viewing directions which are all parallel to the (111) interface. They present different information of the sample and support each other.

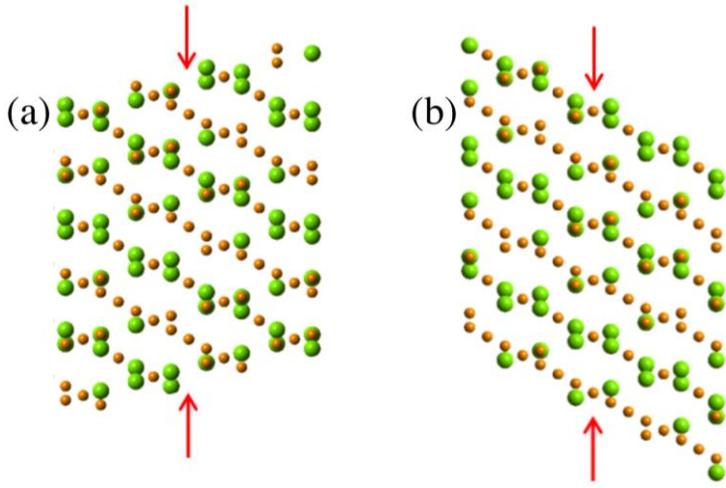

Figure S1. (a) The atomic model of YIG along [$\bar{3}$21] zone axis direction. The oxygen is invisible for clarity. (b) The atomic arrangement model of YIG along [3$\bar{1}\bar{2}$] zone axis direction. The oxygen is invisible for clarity. The red arrows mark the same atom plane of the two zone axis directions.

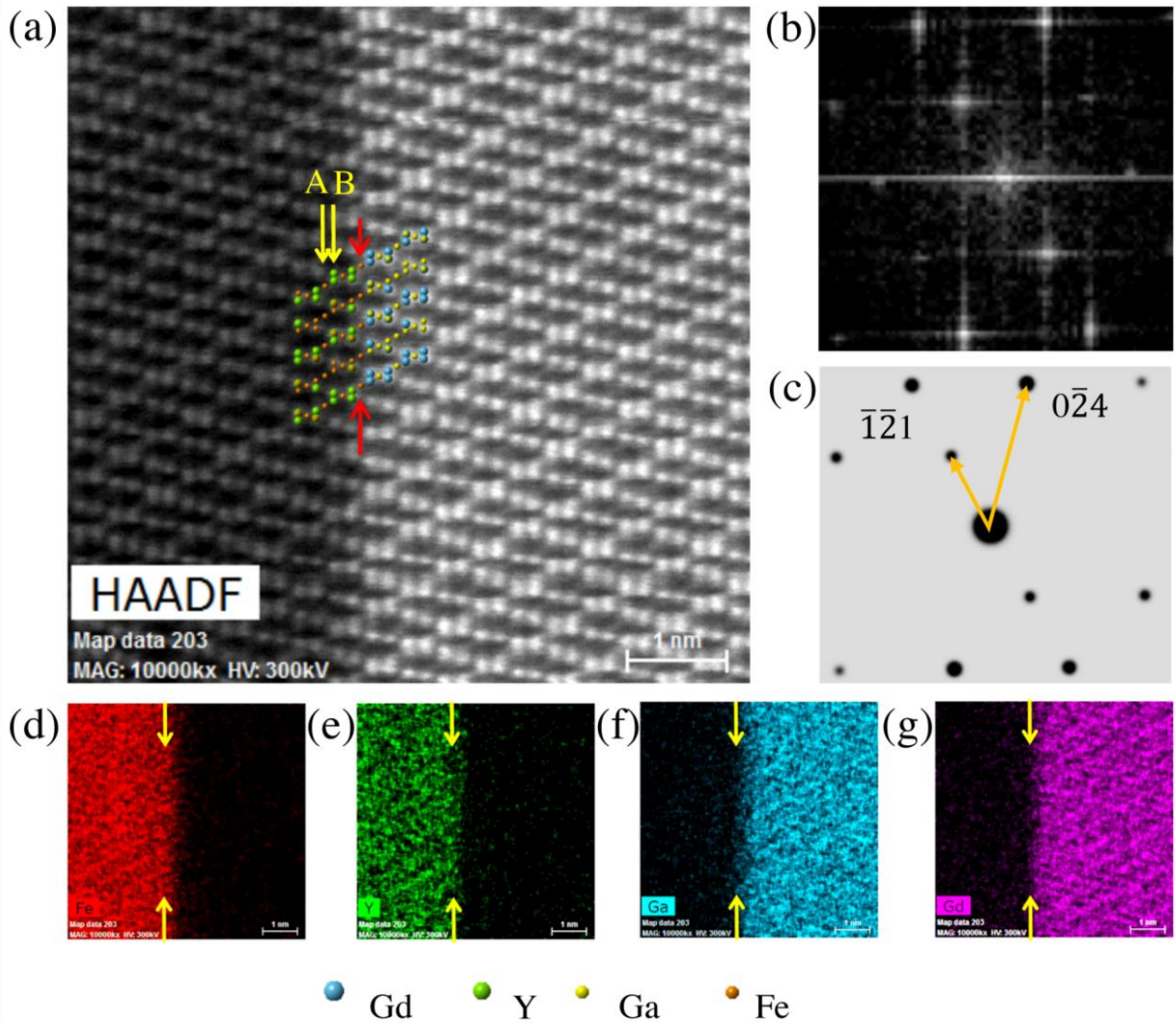

Figure S2. (a) Atomically resolved STEM image with viewing direction of [$\bar{3}$21] direction with atomic model and labeled atom planes on it. The red arrows mark the interface. The yellow arrows mark the two kinds of atom plane of YIG. (b) The Fourier transformation pattern. (c) The simulation of electron diffraction. (d)-(g) EDS maps of (d)element Fe, (e) element Y, (f) element Ga and (g) element Gd. The scale bar is 1 nm. The yellow arrows mark the interface.

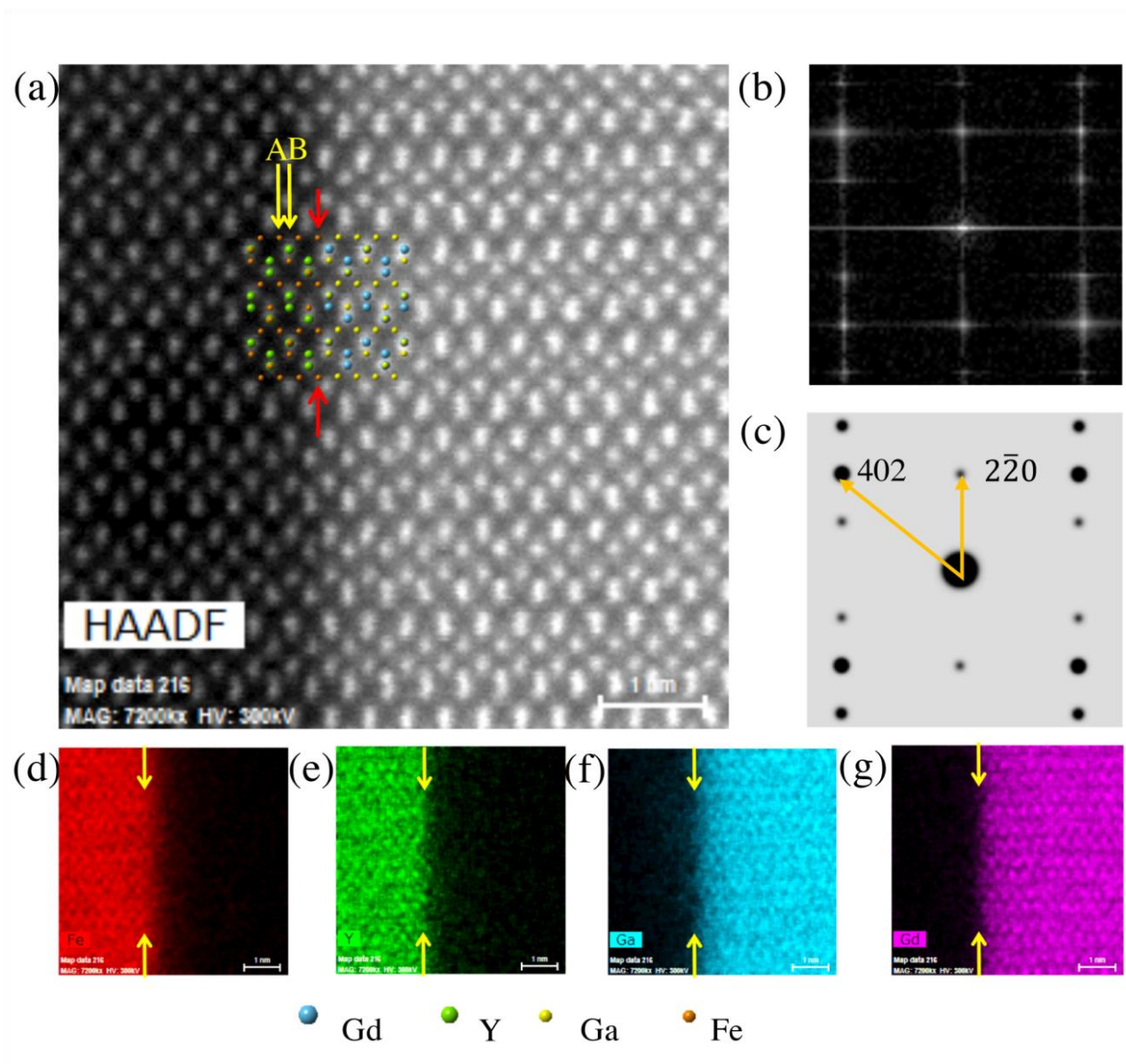

Figure S3. (a) Atomically resolved STEM image with viewing direction of [11$\bar{2}$] direction with atomic model and labeled atom planes on it. The red arrows mark the interface. The yellow arrows mark the two kinds of atom plane of YIG. (b) The Fourier transformation pattern. (c) The simulation of electron diffraction. (d)-(g) EDS maps of (d)element Fe, (e) element Y, (f) element Ga and (f) element Gd. The scale bar is 1 nm. The yellow arrows mark the interface.